\documentclass[conference]{IEEEtran}

\IEEEoverridecommandlockouts

\usepackage{cite}

\usepackage[english]{babel}

\ifCLASSINFOpdf
   \usepackage[pdftex]{graphicx}
\else
   \usepackage[pdftex]{graphicx}
   \DeclareGraphicsExtensions{.eps}
\fi

\usepackage{epstopdf}
\usepackage{url}
\usepackage{enumitem}

\usepackage[final]{changes}

\begin{document}

\title{A Cloud-Based and RESTful Internet of Things Platform to Foster Smart Grid Technologies Integration and Re-Usability}

\author{\IEEEauthorblockN{Alessio Meloni\thanks{This work was supported by Regione Autonoma della Sardegna, L.R. 7/2007: “Promozione della ricerca scientifica e dell' innovazione tecnologica in Sardegna, annualit\`a 2012", CRP-60511. \copyright 2016 IEEE. The IEEE copyright notice applies.} and Luigi Atzori}
\IEEEauthorblockA{DIEE - Department of Electrical and Electronic Engineering\\
University of Cagliari\\
Piazza D'Armi, 09123 Cagliari, Italy\\
Email: \{alessio.meloni\}\{l.atzori\}@diee.unica.it}
}

\maketitle
\begin{abstract}
Currently, one of the hottest topics in the Internet of Things (IoT) research domain regards the issue to overcome the heterogeneity of proprietary technologies and systems so as to enable the integration of applications and devices developed for different environments in a single interoperable framework. Towards this objective, virtualization is widely used to foster integration and creation of new services and applications. Similar benefits are expected by its application in the smart grid arena. The use of a virtualization middleware is expected to enrich the capabilities and opportunities related to the use of smart grid devices. In this paper, we propose key features for an interoperable, re-usable, elastic and secure smart grid architecture which is cloud-based and relies on REST APIs. To highlight the benefits, a practical case study is presented, where outlined functionalities are implemented in a cloud-based IoT platform. The advantages brought by the proposed middleware in terms of interoperability, re-usability, privacy and elasticity of the underlying infrastructure are discussed.
\end{abstract}


\IEEEpeerreviewmaketitle

\section{Introduction}

One of the biggest technological challenges that characterizes major research and development work programmes, such as Horizon 2020 \cite{h2020}, is to overcome the fragmentation of vertically-oriented closed systems, architectures and application areas so as to move towards open systems and platforms that can support multiple applications. The scope is to create an ecosystem composed of devices and services that satisfy crucial requirements such as interoperability, re-usability, scalability and reliability. These requirements are also looked for in the specific context of Smart Grid (SG). Among the key elements of their vision, the Smart Grid's Strategic Research Agenda of the European Union \cite{SGsra2035} and the National Institute of Standards and Technology (NIST) \cite{nist}, identified the need to ensure a successful interfacing of grid equipment between vendors so as to ensure interoperability and integration of the most diverse applications (e.g. Distribution System State Estimation \cite{pau2013} and Smart Home Enegy Management \cite{floris2015}).

As a matter of fact, so far only closed systems relying on a poorly interoperable approach have been used in power systems. Thus, actions such as data collection, storage, access, analysis and visualization have been run independently by involved stakeholders, causing several drawbacks:
 managing resources and data in a closed manner limits new and innovative applications exploiting available resources and integration with cross-domain data (e.g. weather data);
 in a power system with new and updated components such as Distributed Energy Resources (DERs), timely adaptation is compulsory for the correct functioning of the system;
 with the massive amounts of data and information expected to flow with the implementation of SG technologies, data collection and analysis centers become points of criticality in case the necessary infrastructure is not timely scaled;
 since the number of involved components and actors in the SG domain is growing exponentially, the need for a scalable and programmable environment allowing different views of the system is arising, which is currently not satisfied; 
 considering the uncertainty and variability of future power systems, applications need to dynamically adapt their computational power to changes which may occur (e.g. in critical situations).

In light of these considerations, we provide our contribution by proposing a cloud-based virtualization middleware for SG applications so that a digital counterpart of real entities is created allowing for representing components of various nature (smart meters, Phasor Measurement Units (PMUs), concentrators, circuit breakers, servers and many more) into a programmable environment. This proposal can be seen as the first step towards: extending the functionalities, the features and the capabilities offered by entities and devices in various domains; transforming them into services which share functionalities with different stakeholders and give different views depending on permission; composing their capabilities independently from the physical component in order to allow more value-added application innovations. Moreover, virtualizing physical resources \cite{virt_survey} allows for bringing services to the cloud. The use of cloud computing for SG applications \cite{yigit2014} allows for taking advantage of both elastic storage and computational capabilities as well as a cheaper infrastructure from both a CAPital EXpenditure and an OPerating EXpenditure perspective. Furthermore, robustness to outages or failures in a region is provided with \deleted{the }geographic replication\deleted{ of services}. Last but not least, in this paper communication with virtual entities and physical objects take place using a RESTful approach, which guarantees a simple yet highly-performing interfacing between system components \cite{fielding2002}.  

The remainder of the paper is organized as follows. In Section \ref{sec:virt_SG}, the key features of the virtualization framework and the functional blocks of the platform are presented. Section \ref{sec:case_study} presents a case study implemented in a cloud-based Internet of Things (IoT) platform named Lysis \cite{lysis}\added{, used for the management of the interaction among virtual system components and with the physical objects,} in order to show the advantages of the proposed architecture. Section \ref{sec:res} discusses the advantages brought by the proposed architecture. Section \ref{sec:concl} concludes the paper and discusses future work.


\section{Virtualization Framework for the Smart Grid}
\label{sec:virt_SG}

In line with the reference Smart Grid Architectural Model (SGAM) \cite{sgam} developed by CEN, CENELEC and ETSI, and exploiting the \replaced{reference model proposed}{analysis carried out} \deleted{by other works }in the IoT field \cite{icore} \cite{atzori2010}, the \replaced{presented}{proposed} virtualization platform is organized in three levels: 
\begin{itemize}[leftmargin=*]
\item a \textbf{virtualization layer} made of so called Virtual Objects (VOs) that abstract Real World Objects (RWOs) such as Intelligent Electronic Devices (IEDs) \deleted{in the field, station and operation zones, }but also process related components needing some kind of interface to the digital world (e.g. wind turbines), thus hiding the underlying technological heterogeneity of RWOs; 
\item an \textbf{aggregation layer} made of so called Micro Engines (MEs) that are cognitive mash-ups of semantically interoperable VOs created with the aim to aggregate VOs in order to accomplish a certain high-level task and give different views to the application level depending on applications' permissions; 
\item a \textbf{service layer}, which has the role to organize the application requirements into services to be fulfilled thanks to MEs.
\end{itemize}

\begin{figure}[t!]
\centering
\includegraphics [width=1 \columnwidth] {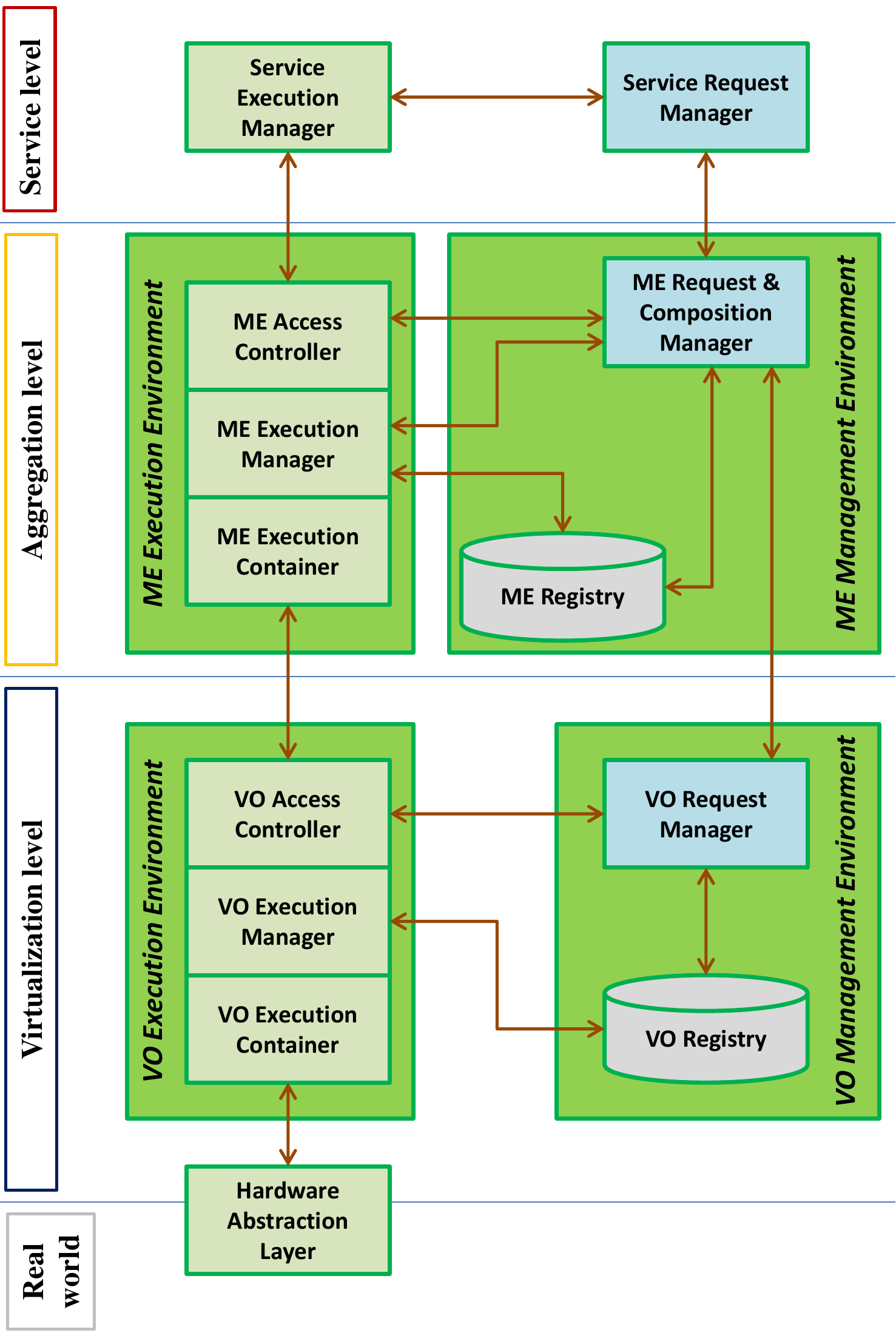}
\caption{Functional blocks of the platform proposed for SG virtualization. Arrows indicate interfaces among blocks.}
\label{architecture}
\end{figure}

The use of such three-layered solution \replaced{is becoming a de-facto reference for the IoT}{has been considered in various works} \cite{giaffreda2013} \cite{gubbi2013}. However, its application to the SG domain has never been fully exploited\added{ by detailing the necessary functional blocks and how the SG functions are mapped to these three layers. As a matter of fact, the SG domain is a complex system of systems with applications of the most diverse nature that need to be coordinated for the sake of interoperability and reusability}. In the following, we introduce different functional blocks composing the overall virtualization framework (Fig. \ref{architecture}) that satisfy the specific requirements of the Smart Grid.


\subsection{Virtualization Layer}

A VO is an abstract description that virtualizes one or more RWOs, so that any application can access it or its functionalities in an interoperable way, without knowing about the means that are needed to physically reach and retrieve information from it. The scope\deleted{ of VOs} is to help building MEs, which comprise one or more VOs, as well as some additional functionalities\deleted{ on top of the VOs}. Thus VOs help in RWO functionalities abstraction through a uniform way of representing data. For this reason, VOs are reusable by different MEs and are permanently and continuously active, monitoring and reacting to the needs coming from the aggregation level as well as to changes in the RWOs. \added[remark=si puo' togliere se si necessita spazio]{As an example, a VO may abstract a PMU and its sensed data such as frequency, current, voltage, thus decoupling these measures from the physical object and describing them by means of metadata such as the timestamp, measurement location and its quality.} 

\added[remark=anche questa parte puo' essere tolta volendo]{Concerning the VO-RWO interface, there are two types of possible interfacing: (a) abstraction of sensed \deleted{RWO }data and metadata coming from the real world; (b) abstraction of the actuation capabilities which influence the \replaced{real world}{RWO}. Each VO can contain either one of both these kinds of interfacing. The role of the VO-RWO interface is to make RWO functionalities generic enough so that either RWO with different functionalities or with similar functionalities but different underlying hardware can interoperate. }This helps the application developer to build aggregated functions very easily. For example, both a Smart Meter and a PMU are able to provide data about the power flowing on a given branch. By means of a virtualization layer, these measures are decoupled from the physical object and presented as data described by metadata such as timestamp and quality of the measure.
\replaced{Last, VOs}{The VO} must also provide a northbound interface to the aggregation level, so that communications with the higher levels of the virtualization platform is put in place. \deleted{In particular, an interface needs to be available between VO and the ME for communicating changes in one level that can have implications on the other level. For example, when a change of status of a RWO is sensed at the virtualization level (e.g. an increased load exceeding a certain threshold), information about this change could be communicated to the aggregation level in order to take proper action. }Herein, all the functional blocks composing the virtualization layer are presented and their role and responsibilities are extensively described.
\begin{itemize}
\item \textbf{Hardware Abstraction Layer (HAL)} is the interface between real world and VOs which can either reside on the physical object as northbound interface, or in the cloud as southbound interface of the VO, depending on the RWO characteristics and capabilities. HAL translates specific vendor functionalities of RWOs into a platform-understandable language. This block is of utmost importance, since it is the only one interfacing with the real world and speaking the RWO specific language. Moreover, it provides a standard communication with the virtualization level by means of REST APIs (which are better explained in the VO Execution Container part), and a standard representation using key-value pairs in JSON format. Doing so, the virtualization platform becomes resilient to components' evolutions, since RWO representation at the virtualization level is standard. 
\item \textbf{VO Execution Environment} represents a virtual environment at the virtualization level, which is most likely owned by the same owner of the physical object, and made of three blocks: 
\begin{itemize}
\item \textbf{VO Execution Manager} is responsible for the management of the VO. It monitors \replaced{its}{the} state\deleted{ of the VO}, QoS parameters as well as connectivity with the RWO and reacts to changes (e.g. a change in the QoS) by updating important information about the VO status in the VO Registry and alerting the ME\deleted{ Execution Manager} if needed. 
\item \textbf{VO Access Controller} deals with MEs permissions to access the VO and manages access conflicts that may arise among MEs\deleted{ trying to act upon a RWO through the VO} (e.g. giving priority to a given ME dealing with safety critical operations over one which optimizes Demand Response).
\item \textbf{VO Execution Container} manages actuation and data retrieval from \replaced{the real world}{RWOs} in terms of which data need to be retrieved as well as how often they need to be retrieved. Depending on the characteristics of RWOs and on the needs, VO can either receive periodic updates from the RWO automatically or request them. The kind of interfacing used depends on the specific RWO and on the HAL. Same considerations can be done for the actuation. \deleted[remark=gia' detto in precedenza per cui volendo si puo' togliere]{Nevertheless, regardless of the underlying interfacing, the VO Execution Container must organize data in a semantically understandable format for the goal of composition at the aggregation level.}
\end{itemize}
\item \textbf{VO Management Environment} is the part of the platform that manages \replaced{all}{the virtualization level and all the} registered VOs. It is made of two components:
\begin{itemize}
\item \textbf{VO Registry} keeps track of the description (VO identifier, location, offered functionalities, etc.) and status of all the running VOs. Therefore, each time the status of the VO changes, the\deleted{ VO} registry is updated by the VO Execution Manager using the HTTP POST method.
\item \textbf{VO Request Manager} is the block dealing with the phase in which requests are received to compose functionalities at the aggregation level. The VO Request Manager is responsible for searching available VOs in the VO Registry based on the requirements coming from the ME Request \& Composition Manager (see next subsection) and the permissions which the requester has with regard to the available VOs, which are granted with a key. Here we suppose three kind of permissions: a VO can be freely accessible if public, accessible only by authorized parties with a \textit{friend key}, accessible only by its owner with a \textit{private key}.\deleted{ Key distribution is thus required only in the case of the friend key.} The VO Request Manager is also responsible for updating the VO Access Controller about new MEs that are granted access to the VO. 
\end{itemize}
\end{itemize}

\subsection{Aggregation Layer}

A ME comprises multiple VOs in order to achieve a specific functionality but it is more than just the aggregation of its parts. While VOs provide a well-defined set of fine-grained functionalities that are abstracted from RWOs and provided for reuse, the task of the aggregation level is to use these fine-grained functionalities and provide higher-level functionalities required by one or more services. Moreover, different MEs may use the same set of VOs to provide different views to the service level which are compliant with stakeholders permissions and ensure proper privacy and confidentiality \cite{erkin2013} \cite{mckenna2012}. For example, if a number of Smart Meters pertaining to a certain zone are considered, a ME can provide consumption data which are aggregated over households but fine-grained over time for the sake of operational purposes. On the other hand, another ME may provide consumption data aggregated over a certain period (e.g. one month) but fine-grained for each household thus providing the necessary information for billing purposes. 

Therefore, the aim of MEs is to effectively link the service level and virtualization level by providing at the same time the glue intelligence and separation needed. Notice that while VOs are entities that represent a fixed and always-on interface to the RWO, MEs can be dynamically modified in order to satisfy service requirements. Here, we define the functional blocks that can be identified at the aggregation level.
\begin{itemize}
\item \textbf{ME Execution Environment} are made of three blocks:
\begin{itemize}
\item \textbf{ME Execution Manager} is the counterpart of the Execution Manager at the VO level. It keeps track of the status of the ME reacting to events that may change MEs status thus needing a specific action. In particular, it monitors requirements dictated by service level and uses self-management functions to meet those requirements and react to changes (e.g. asking the ME Request \& Composition Manager to reconfigure the ME by choosing an alternative set of VOs when service requirements cannot be met).
\item \textbf{ME Access Controller} deals with permissions to access the ME similarly to what has been presented for the virtualization level. Moreover, it has a key role when a ME instance is jointly used by multiple services. In fact, different services may affect the same resources operating simultaneously. This block resolves undesired conflicts and system instabilities. 
\item \textbf{ME Execution Container} is the execution environment of the ME instances in which data coming from the virtualization level are received, parsed and stored or sent to the necessary destination. Moreover, commands coming from the upper layers are passed to the appropriate VO through this block. In other words, it takes care of actuation and data retrieval from VOs, which data need to be retrieved and how often. ME can either receive periodic updates from the VO automatically or request them with a GET HTTP method. In the latter case, the GET method will specify all the parameters necessary for security as well as query related parameters (e.g. for billing purposes, data about household consumption on a given interval). In the former case, data is received automatically in a periodic or triggered-based manner, and the content of the updates as well as its frequency can be modified. In particular, the VO will send an HTTP POST method containing all the \replaced{necessary keys}{parameters which are necessary for authentication and security} as well as commands such as `start periodic updates', `stop periodic updates', `change information of the periodic updates'. Actuation commands take place in a similar manner.
\end{itemize}
\item \textbf{VO Management Environment} is the part of the platform at the aggregation level which is made of two components:
\begin{itemize}
\item \textbf{ME Registry} contains information for each ME such as ME identifier, VOs connected to the ME and many more. Thus the focus of the ME registry is to keep track of all MEs.
\item \textbf{ME Request \& Composition Manager} handles new requests coming from a given Service Manager, verifies the key granting access permission and performs a search in the ME registry in order to find the needed ME. Then, it communicates to the service manager how to reach the considered ME and updates access changes in the ME Access Controller. This block is also responsible for VO composition  changes requested by the ME Execution Manager. 
\end{itemize}
\end{itemize}

\subsection{Service Layer}

The Service Layer consists of functions (i.e. services) that enable SG use cases. A service represents a logical entity which performs a dedicated application task (e.g. visualization of household consumption with a dedicated application).

\begin{itemize}
\item \textbf{Service Request Manager} is responsible for providing requests to the aggregation level in the boot process of a new service.
\item \textbf{Service Execution Manager} is responsible for managing and supervising the execution of the service. 
\end{itemize}

\section{Case study}
\label{sec:case_study}

\begin{figure}[t!]
\centering
\includegraphics [width=0.9 \columnwidth] {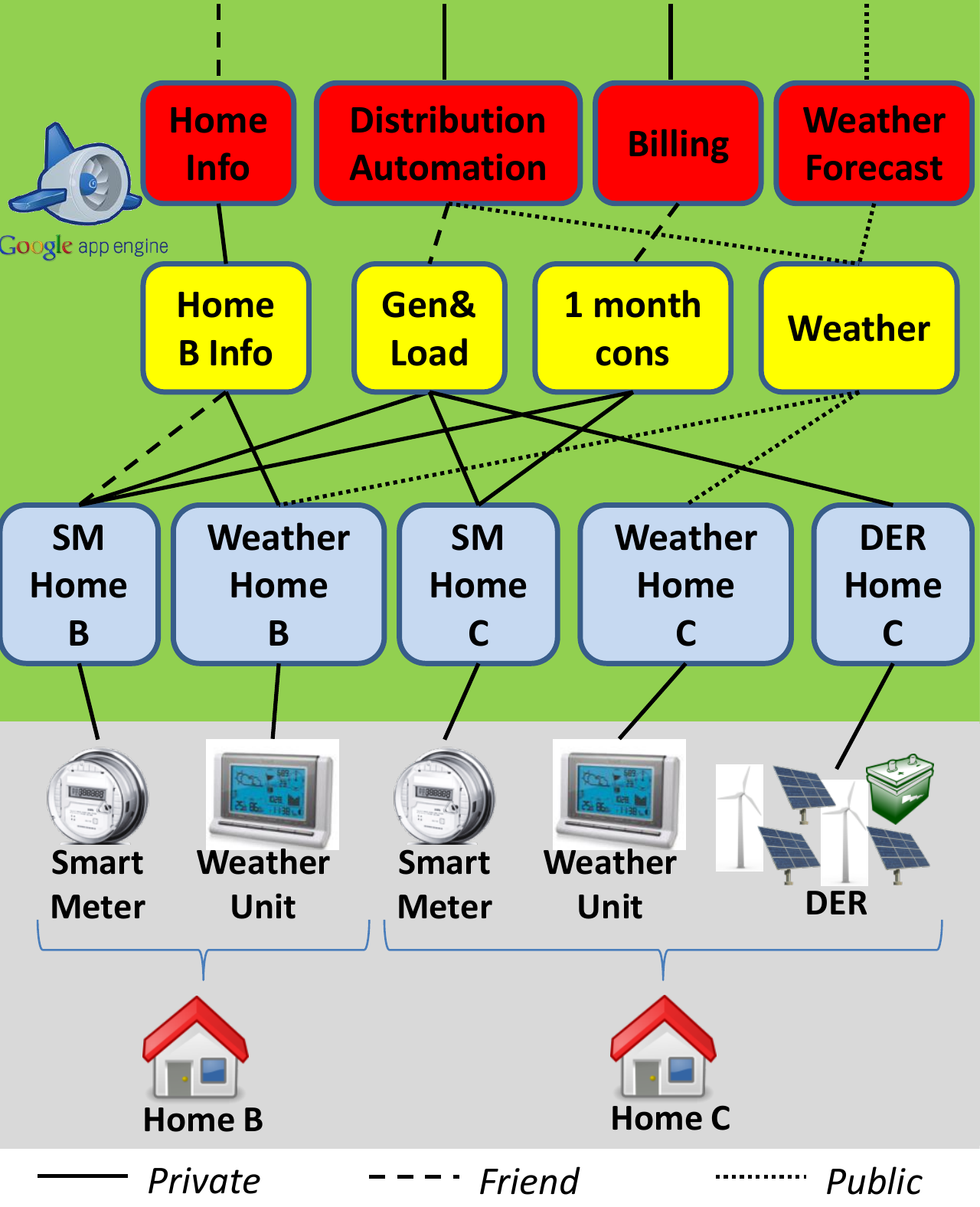}
\caption{Considered case study.}
\label{CaseStudy}
\end{figure}

In this section, the implementation of the proposed architecture is described for the case study of two smart homes, from which data about the consumption, the generation of DER and realtime weather is provided (Fig. \ref{CaseStudy}). For the simulations, we rely on the open UMass Smart* Home Data Set of \cite{barker2012} which is provided in the form of CSV files. In particular, Home B and C of the data set have been considered. For our simulations, these files have been uploaded in 5 different Raspberry Pi 2 Model B development boards
, which can nevertheless be used in a real scenario, considering their capability to implement the needed HAL for objects like microturbines and PV panels, as well as the interface to sensors such as temperature and humidity sensors. This represents our real world level. Concerning the implementation of the virtual entities in the cloud, Google App Engines\added{ (GAEs)} have been used\added{ to host the execution environments of all the three levels (VO Execution Environment, ME Execution Environment, Service Execution Manager)}. \replaced{GAEs}{which} provide a complete Platform as a Service (PaaS) in which the various \added{execution }components have been deployed. \added{The management blocks of all three levels are hosted in the Lysis platform \cite{lysis}, which is also cloud-based and cares about all the management aspects connected to physical and virtual entities (VO Management Environment, ME Management Environment, Service Request Manager).}


\begin{figure}[t!]
\centering
\includegraphics [width=0.9 \columnwidth] {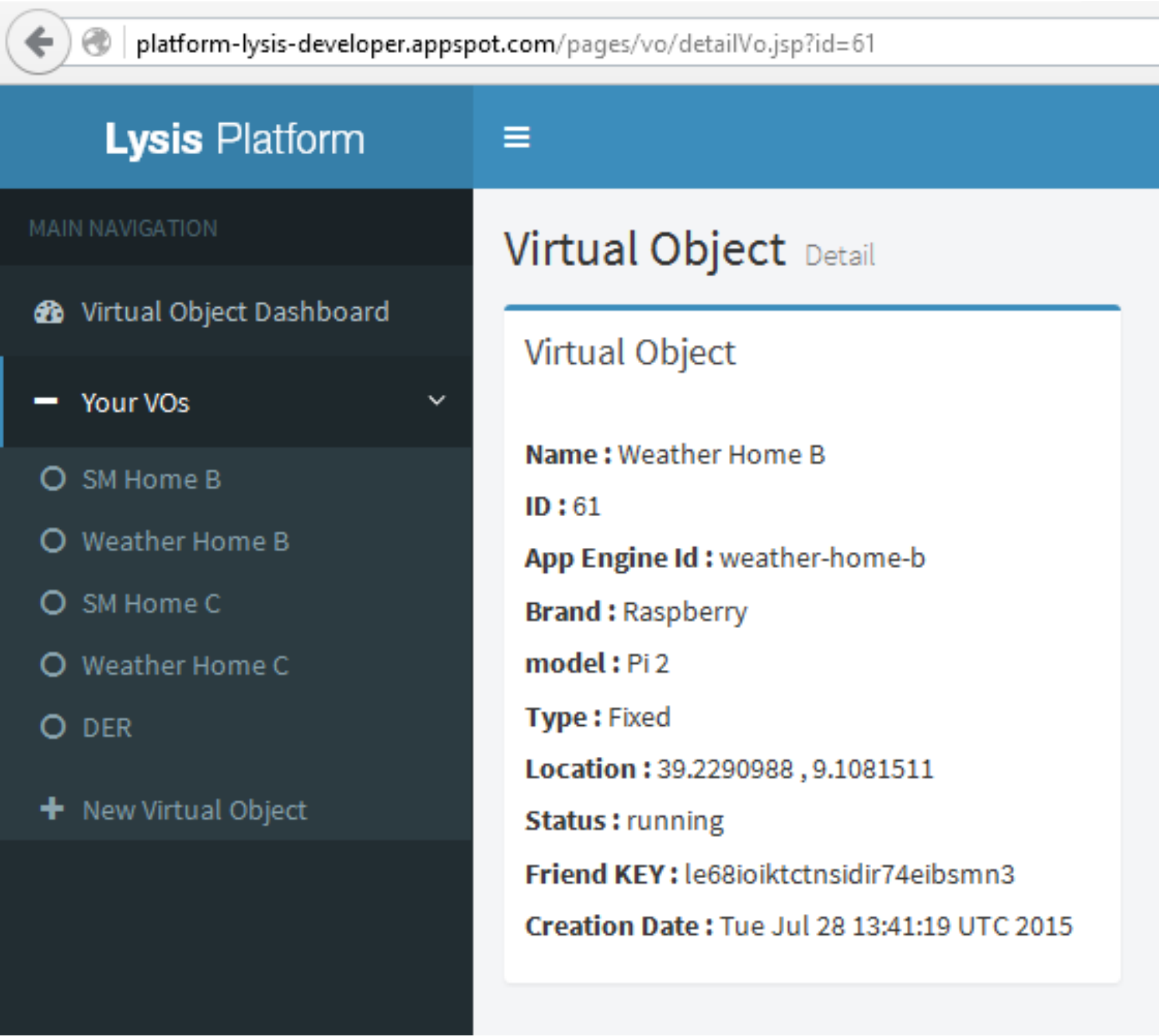}
\caption{A screenshot of registered VOs in the Lysis platform.}
\label{fig:lysis}
\end{figure}

Let us now describe the role of each real and virtual entity, its accessibility (private, friendship-based, public) and its ownership. The next section will leverage on this discussion to highlight the properties claimed in the introduction. Three kinds of RWOs are considered here. Weather units are supposed to be owned by resident of the home in which the unit is installed. Therefore, also the corresponding VOs have the same owner. These RWOs provide the following measurements: timestamp, outside temperature, outside humidity, wind speed, wind direction, wind gust, heat index. In addition, the weather unit in Home B also provides inside temperature and humidity. This data is sent to the corresponding VO every minute. Fig. \ref{fig:lysis} shows how the registered VOs appear in the platform. 

Concerning SM and DER, currently there is a lot of discussion about whether the customer, the utility or a third party should own them. Here, for the sake of generality we consider both of them to be owned by a third party actor. Therefore, also in this case the VOs representing these physical objects have the same owner of the physical object. Every minute, SMs send information about consumption in kWh and the corresponding timestamp. DER is composed of two wind turbines and 3 photovoltaic panels plus a battery. Therefore, each minute information about power generation and about the battery level are sent to the corresponding VO.

Let us now analyze the aggregation and service level. The \emph{Home B Info} is a ME which is owned by dwellers and gathers data about consumption and from the weather unit. Access to the SM is guaranteed by the friend key provided by the metering operator, while Weather Home B is accessed with the private key. In fact, while we suppose weather units to be publicly accessible, the private key allow to receive privacy-critical information such as inside temperature and humidity, which are not provided with the public access. These information are provided in a Home Info visualization service which can be used by dwellers or dwellers' friends which possess a friend key. The \emph{Gen\&Load} and \emph{1 month cons} MEs provide respectively aggregated generation and load information on a 1-minute basis and information about monthly consumption for each customer. These are provided by the third party managing SMs and DERs thus guaranteeing the needed level of privacy. Services such as Distribution Automation of Distribution System Operators and Billing of Energy Retailers can access this information with a friend key and use the gathered information for their own purposes. Finally, public information gathered from the weather unit is aggregated in a \emph{Weather} ME which, for example, could be owned by the municipality in order to provide open data for local weather forecast services or for distribution automation services which exploit weather conditions.  

\begin{figure}[t!]
\centering
\includegraphics [width=1 \columnwidth] {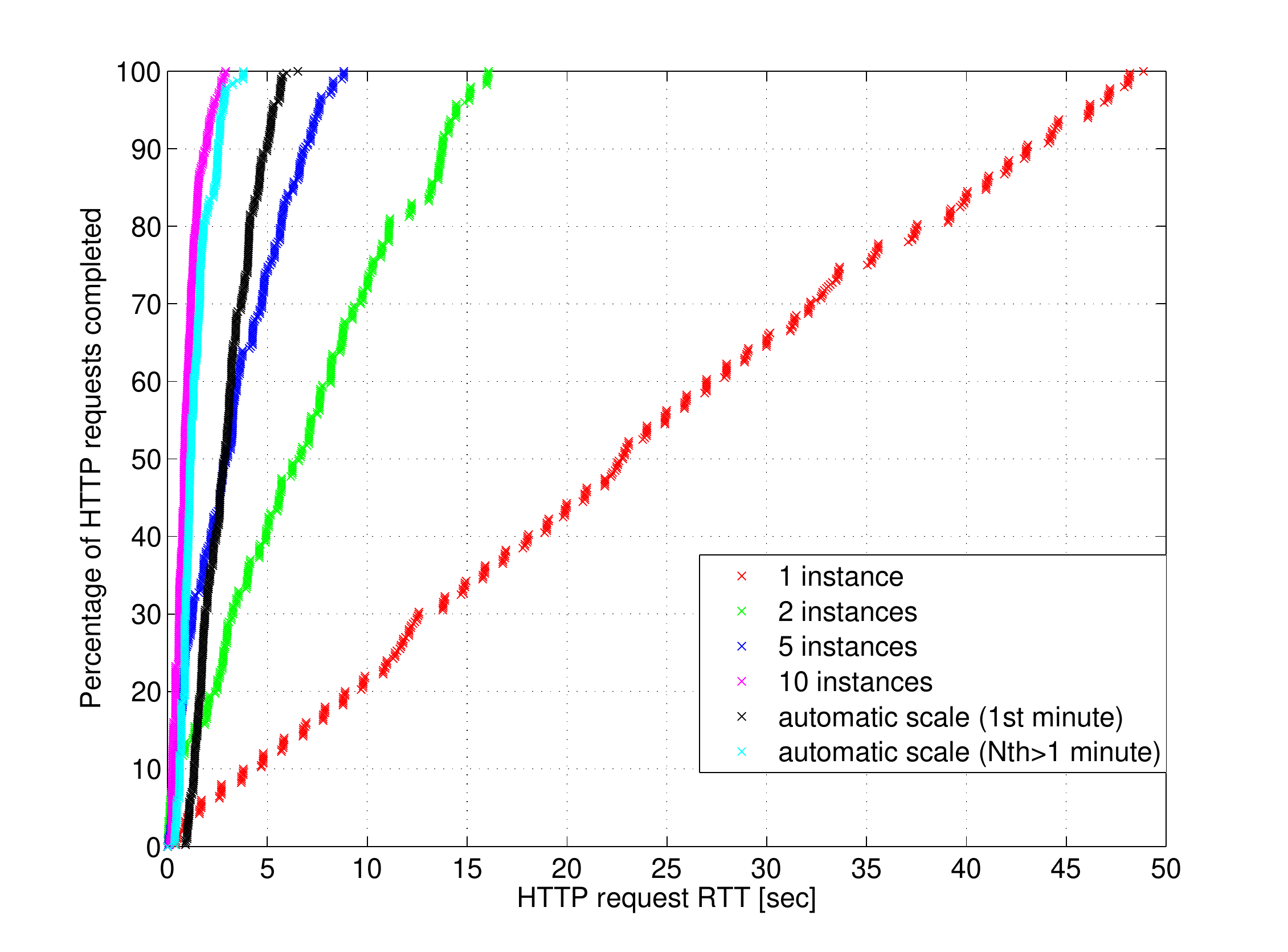}
\caption{Google App Engine promptness to HTTP requests depending on instances' settings.}
\label{fig:microgrid}
\end{figure}

To demonstrate the adaptability property of the cloud, the UMass Smart* Microgrid Data Set of \cite{barker2012} has been used, which comprises SM consumption of 400 homes in the same period. Here, we assume all 400 SMs sending an HTTP post to the virtual entity in the cloud, at the beginning of a new minute. Fig. \ref{fig:microgrid} shows how the performance changes with the number of instances. Instances are \replaced{clones}{deployment copies} of the virtual entity which can be created in order to satisfy high loads\added{ of HTTP requests by parallel processing}. This process is not visible to the senders, which utilize the same address\added{ and are not aware of the presence of more than one instance per virtual entity}. The number of instances can be statically or dynamically set. If dynamically set, parameters such as the minimum and maximum number of instances and triggers for scaling up and down the number of instances must be specified. From Fig. \ref{fig:microgrid} it can be seen that the bigger the number of instances, the better the performance of the virtual entity in satisfying HTTP requests. In our particular case study this is important, since data for operational purposes must be updated each minute and using an instance means to delay data acquisition by almost $50$ seconds. Finally, we have also simulated the case in which instances are dynamically scaled in order to reduce costs in a pay per use PaaS. In particular, instances were allowed to scale from a minimum of zero (idle) to a maximum of 10. Results show that at boot (1st minute) the performance is initially bad but rapidly improves as soon as the instances are created. On the other hand, in all the following minutes ($N>1$) the performance is comparable to the one for the same static number of instances.

\section{Results Analysis}
\label{sec:res}

In this section, the properties which result from the deployment of the considered virtualization middleware in the considered case study are analyzed. 
\subsubsection{Reusability}
In the proposed case study, weather data is of big interest for: the owner of the house, who would like to know the weather at her dwelling location; the distribution operator, which will combine this information with energy generation and consumption for distribution optimization; the municipality, which can exploit various sensors across the city in order to provide open data. Therefore, various actors can benefit from the same physical objects with just one of them needing to install and maintain sensors.  
\subsubsection{Integration}
The two weather stations in Home B and C have different properties, since only the first one can provide also data about indoor temperature and humidity. Nevertheless, these differences are not perceived on the aggregation level, since the particular properties of each RWO are decoupled at the virtualization level. But integration also means putting together data from different domains such as weather and energy consumption, which boosts the creation of new services such as distribution automation.
\subsubsection{Adaptability}
Different services have really diverse and sometimes variable computational needs. Leveraging on cloud services such as Google PaaS allows to overcome these problems in an affordable way, since resources can be easily scaled and paid per use.
\subsubsection{Security \& Privacy}
The use of keys to access data allows for reusing data coming from the same RWO without compromising privacy. In our case study, the SM can be used both for network operational management and for billing purposes. As a matter of fact, by using access keys and MEs, it is ensured that actors will gather data with no more than the granularity (in terms of time or in terms of SMs data aggregation) which has been granted to them.\added{ Therefore, users' concerns on information leaking such as non-intrusive load monitoring and operators concerns related to the economic value are properly managed.}

\section{Conclusions and Future Work}
\label{sec:concl}

In this paper an Internet of Things platform which is cloud-based and relies on a RESTful interfacing has been presented. This middleware allows to break vertical domain silos of individual applications, thus making SG an integral part of the Internet of Things. \replaced{The }{In particular, leveraging on the recently created Lysis platform, the }presented architecture satisfies compulsory requirements for Smart Grid deployment such as reusability, integration, adaptability, security and privacy. The core rationale of this paper is to present the concept and functional blocks behind the idea of a virtualization platform with RESTful interfacing. The scenario presented completes the core idea by giving a practical case study in which it can be used\added{ for SG applications with very different requirements and exploiting cross-domain data to further optimize application-goals}. Details about APIs have been voluntarily omitted for the sake of brevity. In a extended version of this paper, we would like to present more application scenarios and give extensive information about how to create VOs starting from devices with REST capabilities and about APIs which are used to access VOs we created. The goal is to realize a set of VOs that can be used and exploited by anyone interested, in order to advance research. 


\end{document}